\documentclass[table]{bmcart}

\usepackage[utf8]{inputenc} 

\usepackage{amssymb}
\usepackage{latexsym}
\usepackage[cmex10]{amsmath}
\usepackage{framed,multirow}

\usepackage{graphicx}
\usepackage[caption=false]{subfig}

\usepackage{url}
\usepackage{xcolor}
\usepackage{colortbl}

\definecolor{lightergrey}{RGB}{240,240,240}
\definecolor{lightgrey}{RGB}{210,210,210}
\definecolor{grey}{RGB}{150,150,150}
\definecolor{darkergrey}{RGB}{100,100,100}

\usepackage{enumitem}

\usepackage{hyperref}

\usepackage{hhline}
\usepackage{array}

\newcolumntype{P}{>{\centering\arraybackslash}p{5.1em}}
\newcolumntype{B}{>{\centering\arraybackslash}p{1.25cm}}
\newcolumntype{C}{>{\centering\arraybackslash}p{6em}}

\begin{document}

\begin{frontmatter}

\begin{fmbox}
\dochead{Research}


\title{Automatic detection of alarm sounds in a noisy hospital environment using model and non-model based approaches}


\author[
   addressref={aff1},                   
   corref={aff1},                       
   noteref={n1},                        
   email={ganna.raboshchuk@upc.edu}   
]{\inits{G}\fnm{Ganna} \snm{Raboshchuk}}
\author[
   addressref={aff1},
]{\inits{S}\fnm{Sergi} \snm{G\'{o}mez Quintana}}
\author[
   addressref={aff1},
]{\inits{A}\fnm{Alex} \snm{Peir\'{o} Lilja}}
\author[
   addressref={aff1},
]{\inits{C}\fnm{Climent} \snm{Nadeu}}


\address[id=aff1]{
  \orgname{TALP Research Center, Department of Signal Theory and Communications, Universitat Polit\`{e}cnica de Catalunya}, 
  \city{Barcelona},                              
  \cny{Spain}                                    
}


\begin{artnotes}
\note[id=n1]{Corresponding author} 
\end{artnotes}

\end{fmbox}


\begin{abstractbox}

\begin{abstract} 
In the noisy acoustic environment of a Neonatal Intensive Care Unit (NICU) there is a variety of alarms, 
which are frequently triggered by the biomedical equipment. 
In this paper different approaches for automatic detection of 
those sound alarms are presented and compared: 1) a non-model-based approach that employs signal processing techniques; 
2) a model-based approach based on neural networks; and 3) an approach that combines both non-model and model-based approaches. 
The performance of the developed detection systems that follow each of those approaches is assessed, 
analysed and compared both at the frame level and at the event level by using an audio database recorded in a real-world hospital environment.
\end{abstract}




\end{abstractbox}
%

\end{frontmatter}



\section{Introduction}

Very low birth weight preterm infants often have health problems and
receive a specialised medical care in a Neonatal Intensive Care Unit (NICU)
during the first several weeks or even months of life, what is crucial for their survival.
In the acoustic environment of a typical NICU there is a diversity of sounds, produced either by the human activities or by the biomedical equipment
\cite{hassanein_noise_2013,livera_spectral_2008}, which often happen simultaneously and
contribute to excessively high sound pressure levels \cite{krueger_levels_2005}.
The possible harmful effects of such a noisy environment on the further growth
and neurodevelopment of the preterm infants have been well documented and are of great concern in the medical literature
\cite{wachman_effects_2010}.

Acoustic alarms generated by various types of biomedical equipment are extensively present in a NICU environment
and are used to alert of situations requiring medical attention.
The fact that a large number of sounding alarms are not clinically relevant and/or are unrelated to emergency situations
\cite{freudenthal_quiet_2013} may cause alarm fatigue and lead to
unsatisfactory quality of healthcare provided by the medical staff.
Intelligent alarming systems, which make use of alternative alarm modalities and keep only the most critical alarms sounding,
are being proposed in order to improve the alarm handling process in NICUs and to reduce noise levels \cite{freudenthal_quiet_2013,van_pul_alarms_2015}.
Unfortunately, in the majority of the NICUs such systems are yet to be developed.

The automatic detection of alarm sounds in a NICU can be useful in two ways.

First, for detecting the sounds that are potentially harmful for a preterm baby due to their
specific spectro-temporal structure. The effects of a NICU acoustic environment
on a preterm infant could be revealed by the infant reactions to auditory stimuli from it,
which can be investigated by relating the presence of particular sounds
with the preterm physiological variables.
Such investigation can complement greatly the work already reported in the literature,
in which only the sound pressure level is considered \cite{kuhn_infants_2012},
and requires big amounts of labelled audio data,
which can hardly be obtained without using automatic detection from audio signals.

Second, for assisting the medical staff in their work and facilitate the reaction to events.
E.g. in \cite{jousselme_sensor_2011} a sound-activated light device was implemented for alerting the
staff members when the sound pressure level exceeded a predefined threshold.
The automatic alarm detection system can be a part of a more sophisticated notification system
allowing smart alarm handling algorithms, which could be designed to warn about triggering of
particular alarms, to take into account their clinical relevance and urgency, etc.

The acoustic alarm detection was previously investigated for the purposes of
hearing impaired assistance or hearing support in noisy conditions~\cite{beritelli_emergency_2006,carbonneau_in_ear_2013}.
To our knowledge, research on the topic was first reported in~\cite{ellis_alarms_2001},
where the detection of various real-world alarm sounds was addressed.
In that work, two different approaches were presented:
a generic model-based approach that employs features capturing the global properties of the spectrum and neural networks,
and a non-model-based signal-separation approach that exploits the specific time-frequency structure of alarms.
While the model-based approach is also followed in~\cite{beritelli_emergency_2006},
most of the posterior works adhere to the non-model-based approach
\cite{carbonneau_in_ear_2013,lutfi_automated_2012,meucci_realtime_2008,xiao_cockpit_2009}.
Differently, in~\cite{schroder_parts_2013} the acoustic siren detection problem is tackled
from the image processing perspective. In that work, the spectrogram is treated as an image and part-based models, which
consist of spectro-temporal patches in relative and flexible time-frequency configurations, are learnt.

The non-model-based systems make use of the particular properties of alarms
in one form or another to provide decisions using rules,
and their performance depends strongly on the proper choice of the threshold values.
The model-based systems perform statistical modelling from a multitude of training alarm
samples in multiple conditions, and the amount of training data usually plays an important role.

In the work reported in this paper we approach the problem of alarm detection from three different perspectives.
From them, several systems were developed and they were compared using a database collected in a real-world NICU environment.

First, a non-model based system which employs matched filter and morphological operators is presented.
The first stage of the system performs the input signal enhancement, which plays a crucial role for the system performance.

Second, two model-based systems are proposed, which use either generic or class-specific spectral inputs to
neural networks \cite{peiro_alarms_2017}. Neural networks were also employed in~\cite{ellis_alarms_2001,beritelli_emergency_2006} using a
conventional topology, and an input consisting of extracted features that cover the entire frequency bandwidth
(i.e. perceptual linear prediction cepstral coefficients in~\cite{ellis_alarms_2001} and
mel-frequency cepstral coefficients in~\cite{beritelli_emergency_2006}).
Conversely, in our work, a simple magnitude spectrum representation
which provides higher frequency resolution is employed at the input, and
partially connected layers for weighting the input data in frequency and in time are explored.

Third, a system that represents a combination of the non-model and model-based approaches
and takes advantage of both of them. It performs statistical modelling of the training data, but also uses
the knowledge about spectral and temporal alarm characteristics. The spectral knowledge is used to form the feature vector,
and the temporal knowledge is incorporated at a post-processing stage.
This system was initially reported in~\cite{raboshchuk_alarms_2015}.
The non-model-based system that exploits the knowledge about
the alarm characteristics in a similar manner was reported in~\cite{xiao_cockpit_2009}.
Unlike in that work, our system takes into account: 1) frequency and duration variation observed in the alarms;
2) amplitude structure of the alarms which is important for discriminating the alarms that share specific frequencies.

The rest of the paper is organised as follows. The evaluation setup used to assess
the performance of the detection systems is provided in Section~\ref{sec:AlarmEvaluation},
which includes a brief description of the audio database, the alarm classes, and the evaluation metrics.
General considerations about the development of the detection systems are outlined in Section~\ref{sec:GeneralNotes}.
Sections~\ref{sec:MatchedFilter},~\ref{sec:KnowledgeBased} and~\ref{sec:DeepLearning} describe,
respectively, the non-model-based approach, the model-based approach and the combined approach.
Finally, Section~\ref{sec:Discussion} summarizes, compares and discusses the results obtained by
the various systems that were developed by following those approaches.

\section{Evaluation setup}
\label{sec:AlarmEvaluation}

\subsection{Audio database and alarm sound classes}
\label{sec:Data}

The audio database used in this work contains ten recording sessions,
which were carried out in the NICU of Hospital Sant Joan de D\'{e}u Barcelona.
Each recording session contains a subset of the defined acoustic scenarios, which mostly correspond to the
daily nursery care related activities. The recordings were made using two electret
unidirectional microphones connected to a linear PCM recorder.
One microphone was placed inside the incubator, close to the infant's ear,
and the other one outside the incubator, at approximately 50~cm distance above it,
usually pointing to the centre of the room.
The recordings were downsampled from original 44.1 to 24~kHz.
More information about the database acquisition and the acoustic description of the NICU
can be found in \cite{raboshchuk_alarms_2014}.

\begin{figure}[!t]
\centering
\includegraphics[width=0.7\linewidth]{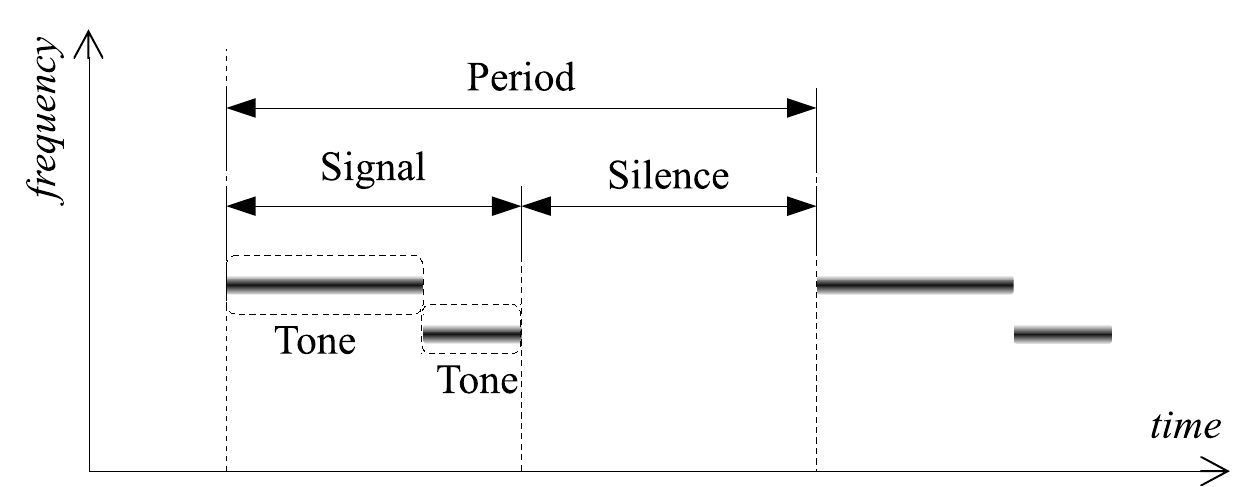}
\caption{General structure of alarm sounds.
Only fundamental frequency is depicted for clarity of presentation.}
\label{fig:Terminology}
\end{figure}

By observing the recorded audio data we found 16 different types of alarms generated by
various biomedical equipment. Only 7 types (classes) were selected for our tests under
the criteria of being the most represented in the database and being relevant from the medical point of view.
These classes are a1, a3, a6, a7, a8, a10 and a16.

The general structure of the observed alarms is shown in Figure~\ref{fig:Terminology}.
The alarms are periodic in time, and each alarm period consists of signal and silence intervals of established durations.
For the selected alarms the signal interval usually consists of a single stationary signal (tone), and only for one
class it consists of two different consecutive tones (as depicted in Figure~\ref{fig:Terminology}).
Each tone contains one or several simultaneous frequency components, which may or may not be harmonically related.

The particular properties of the selected alarms (alarm-specific frequencies, signal and silence interval durations) were carefully analysed.
Three of the alarm classes (namely, a1, a8 and a16) have very similar spectro-temporal properties, i.e. they share some of their class-specific frequencies.
Four of the alarm classes (a1, a3, a7 and a10) show some variation in the frequency and duration values among different device units of the
same model. Since such differences are not perceived by the medical staff, they are referred to as different \textit{versions} of the alarm class.

The manual annotations cover 54.3~min of the acquired audio data, and alarm sounds occur 19.28\% of this time.
Note that each alarm signal, the non-silence interval of each alarm period (see Figure~\ref{fig:Terminology}), was annotated separately.
The duration of audio segments annotated as a specific alarm signal varied from 1.24 to 5.02\% of the total data duration,
and two or more alarm signals occurred simultaneously during almost 8\% of it.

\subsection{Evaluation metrics}

The performance of the detection systems was assessed both at the frame level and at the event level.
For the frame-level evaluations, the Missing Rate (MR) and the False Alarm Rate (FAR) metrics were used, which are defined as
\begin{equation}
  MR=\frac{N_{M}}{N_{A}}, \hspace{5mm} FAR=\frac{N_{FA}}{N_{NA}},
\label{Eq:FBmetrics}
\end{equation}
where $N_{M}$ and $N_{FA}$ are the number of misclassified frames for alarm and non-alarm class, respectively,
and $N_{A}$ and $N_{NA}$ are the total number of alarm and non-alarm frames, respectively.

In addition, we also propose an event-level evaluation metric that is more meaningful for the medical application.
The period of the alarm is chosen as event, so each alarm period is a different event.
The Period-Based ERror Rate (PB-ERR) is defined as complementary of $F_1$-score as
\begin{equation}
\textit{PB-ERR} = 1 - F_1 = 1 - \frac{2 \cdot N_{C}}{2 \cdot N_{C} + N_{FA} + N_{M}},
\end{equation}
where $N_{C}$ is the number of correctly detected alarm periods.
Each reference period is regarded as correctly detected if there exists a detected alarm period
in the tolerance interval $[T_{ref}-T_{tol}; T_{ref}+T_{tol}]$,
where $T_{ref}$ is the reference period beginning timestamp and $T_{tol}$ is the tolerance interval duration,
which was set to $49\%$ of the alarm period duration.
In this case, the system is expected to detect an alarm in the tolerance interval that has the duration of almost one alarm period.

A 10-fold cross-validation scheme was applied in order to obtain more statistically relevant results. On each fold,
9 recording sessions were used for training and 1 session for testing.
The overall metric scores were obtained by aggregating the results over all 10 folds.
The reported results correspond to the average of class-based metric scores.

\section{Development of the detection systems}
\label{sec:GeneralNotes}

In this work, three different systems were developed: a non-model-based system, a model-based system, and a system that combines both non-model and model-based approaches.
Each detection system consists of a set of binary detectors, where an individual detector
is designed to deal with a particular alarm class and is trained following the one-against-all strategy (alarm class vs. non-alarm class).
The input audio is processed by each binary detector independently. That approach is adopted due to the following reasons:

\begin{itemize}[noitemsep,topsep=0pt]
 \item A set of considered alarm classes often is not definitive.
 Building individual binary detectors provides flexibility as the detection system can be easily extended to new classes.
 \item A detector designed for a particular alarm class can exploit its specific properties to achieve a better discrimination.
 \item In case of temporal overlaps between alarm sounds, multiple output labels can be provided 
 for the audio region with overlapping.
  \item Already developed individual detectors could be reused in another NICU room for coinciding alarm classes.
\end{itemize}

Due to their specific characteristics, the model-based system performs classification at the frame level, while
the non-model-based system performs it at the the alarm period level.
The combined system initially provides decisions at the frame level,
but includes a post-processing scheme (with temporal modelling, see Section~\ref{sec:PostProcessing})
for obtaining period-level decisions.
For systems operating at the frame level, the decision threshold is chosen based on the Equal Error Rate (EER) criterion,
so assuming that both miss and false alarm errors are equally important at the frame level.
In these systems, for the period-level evaluation, the beginning of a sequence of consecutive frames of any length
detected as belonging to the alarm class is regarded as the timestamp of the alarm period label.
For the systems operating at the period level, to obtain the frame-level decisions, a number of frames corresponding to
the duration of the signal interval are assigned to the alarm class at each of the detected alarm period timestamps.
In all cases, a constraint of minimal time distance between consecutive detected periods is applied,
where the minimal distance is taken equal to 75\% of the alarm period duration.

\section{Non-model-based approach}
\label{sec:MatchedFilter}

\begin{figure}[!t]
\centering
\includegraphics[width=0.9\linewidth]{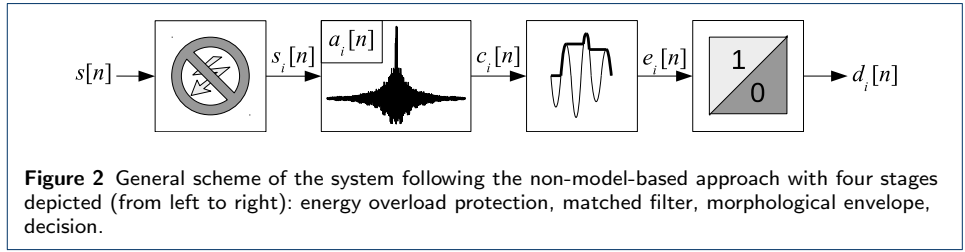}
\caption{General scheme of the system following the non-model-based approach with four stages depicted (from left to right):
energy overload protection, matched filter, morphological envelope, decision.}
\label{fig:System_scheme}
\end{figure}

The proposed system is based on matched filter and morphological operators, and consists of 4 different stages (see Figure~\ref{fig:System_scheme}).
The first is the Energy Overload Protection (EOP) stage, which performs a prior enhancement of the input signal $s[n]$.
The second stage, Matched Filter (MF), which computes the correlation between its input signal $s_{i}[n]$
and the reference sample of alarm signal of class $i$, $a_{i}[-n]$. 
At the third stage that sequence $c_{i}[n]$ is processed with morphological operators to obtain an envelope $e_{i}[n]$.
Finally, at the last stage a binary decision $d_{i}[n]$ about whether the alarm is detected or not is taken.
In the following subsections these stages are described in detail.
The EOP stage was added posteriorly for improving the detection results and therefore it is the last to be described.

\subsection{Matched filter}

At this stage, the cross-correlation between the input signal $s_i[n]$ and the reference alarm signal $a_i[n]$
of duration $L$ is computed as:
\begin{equation}
c_i[n] = \frac{\sum_{k=0}^{L-1} s_i[n+k]a_i[k]}{\sum_{k=0}^{L-1}a_i^2[k]}.
\label{Eq:MF_output}
\end{equation}
Note that for some alarm classes the detector contains a bank of MFs,
each dealing with a different version of that alarm class.

Since the recorded alarm reference may have a noise floor at non-alarm frequencies,
it is filtered to obtain a clean reference signal. 
The filter removes the spectral content out of a set of relevant frequencies, 
which are estimated from the noisy alarm reference signal.
Relevant frequencies are defined as frequencies corresponding to local maxima of
power spectral density such that the quotient between their power and the power of the strongest frequency
is below a threshold value.
Note that that value has to be less than the Signal-to-Noise Ratio (SNR) of the noisy reference, and in this work it is set to 20~dB.
For the alarm with several tones, the relevant harmonics are defined in frames (200~ms, half-overlapped),
where the power spectral density of each frame is normalised with respect to the maximum among all frames.

\subsection{Morphological envelope}

The Morphological Envelope (ME) stage is a concatenation of full-wave rectifier and morphological closing.
Closing is a morphological operator that uses a binary sequence $b[n]$ called structuring element.
In particular, a flat structuring element of size $S$ is used:
\begin{equation}
b[n] = \begin{cases}
        0, & \text{if } 0 \leq n < S \\
        -\infty, & \text{otherwise}
       \end{cases}
\label{Eq:SE}
\end{equation}
where the value of $S$ is based on the fundamental frequency of alarm $f_0$ and is equal to the integer closest to $1/f_0$.
Since closing only takes into account positive peaks, a full-wave rectifier is used previously to obtain absolute values.
For each scenario, the amplitude of the smoothed envelope $e_i[n]$ is normalised by its mode value.

\subsection{Decision stage}

At this stage, the decision about whether each peak at the output of the ME stage corresponds
to an actual alarm period or not is taken.
First, a low-pass finite impulse response filter is used to smooth the envelope signal~$e_i[n]$.
Again, the matched filter concept is used and the filter impulse response is obtained from
the output of the ME stage when the reference alarm signal is introduced at the input of the system.
Further, since the envelope can contain peaks that do not correspond to an actual alarm period,
the values of the envelope below a predefined threshold $U$ are assigned to zero.
The particular value of the threshold is equal to the maximum value of those non-alarm peaks.

\subsection{Energy overload protection}

The recordings often have some strong knocks and glitches, 
which may affect the response of the MF.
In order to deal with that, an EOP stage was added, which consists of a filter followed by a dynamic compressor.
The same filter designed for obtaining the clean reference alarm signal is employed.
A compressor is defined by a threshold $T$ (in dB) and by a compression ratio $R:1$ (in our case, $R=10$).
The compression starts after $T$ is reached.
In this work, the threshold $T$ is estimated statistically from the training data as 90\textsuperscript{th} percentile
(thus, only 10\% of the input signal is compressed).
The input signal is rectified and smoothed before compression by convolving its absolute value with a window.
That window is defined by 3 time parameters:
attack (also known as a look ahead time, 5~ms), sustain (10~ms) and release (50~ms) times,
which are defined based on the length of glitches.

\subsection{Experimental results}

\begin{table}[!t]
\small
\caption{Alarm detection performance obtained by the non-model-based system}
\label{tab:SergiResults}
\vspace{2mm}
\centerline{
\begin{tabular}{p{1cm}p{1.4cm}ccc}
\hline 
\multicolumn{2}{c}{\multirow{2}{*}{System setup}} & \multicolumn{3}{c}{Evaluation metrics (\%)} \\ 
               &  & MR  & FAR  & PB-ERR \\
\hline \hline
\multirow{2}{*}{``Oracle''} & w/o EOP & \textbf{18.50} & 67.99 & 92.46 \\
& w EOP & 27.50 & \textbf{0.19} & \textbf{12.15} \\
\hline
\multicolumn{2}{l}{\textbf{10-fold CV}} & 42.38 & 0.53 & 34.24 \\
\hline
\end{tabular}}
\end{table}

Two different setups were used for the system evaluation.
In the first setup, which we call \textit{``oracle''}, the knowledge about the temporal location of non-alarm intervals (i.e. labels)
is used to estimate the decision threshold $U$ for each scenario sample at the input of the system.
This setup was used during the system development and gives an idea of the upper bound of its performance.
The first part of Table~\ref{tab:SergiResults} presents results for this evaluation setup,
both for the initially developed system without the EOP stage (row 1) and for the finally proposed system with EOP as depicted in Figure~\ref{fig:System_scheme} (row 2).
It can be seen that once the EOP is added as the first stage of the system,
the amount of false alarm errors is reduced drastically,
which leads to a very low PB-ERR metric score of 12.15\%.

In the second setup, \textit{10-fold cross-validation (CV)},
the threshold $U$ is estimated based on the scenarios from training sessions.
For each training scenario a threshold is computed using labels, as was done in the previous setup.
The testing threshold is computed based on those training thresholds as the average between their minimum and maximum values.
Note that the testing threshold is the same for all scenarios of the testing session.
Since the amplitude of the smoothed envelope may vary from session to session,
for each scenario it is normalised by its mode value.

Indeed, the results for the \textit{10-fold CV} evaluation setup (last row in Table~\ref{tab:SergiResults})
are significantly worse than the ones obtained from the \textit{``oracle''} setup, as the threshold~$U$ estimate is less accurate due to mismatched conditions. 
Nevertheless, this setup is more realistic than the \textit{``oracle''} one, and it is used for the comparison with the other detection systems
in this paper.

By analysing the detection results for each alarm class separately, we observed that
the three alarm classes that have similar spectro-temporal properties show the highest scores in terms of MR metric.
This is in agreement with the highest maxima of the normalised cross-correlations between the alarm reference signals.
Basically, a large number of miss errors is caused by the high threshold $U$ estimate,
that results from the high non-alarm peaks obtained from similar alarms present in the training data.

\section{Model-based approach}
\label{sec:DeepLearning}

The two systems proposed in this section are entirely based on the use of Neural Networks (NNs)
and have either a \textit{generic} or a \textit{class-specific} input,
which is based on a raw spectral magnitude.
For that purpose, partially connected hidden layers with limited weight sharing are explored
for weighting the input data in frequency or in time,
which was inspired by recent works reported in \cite{sainath_learning_2013,abdelhamid_exploring_2013}.
Due to the limited amount of available annotated data, the employed network structures are small size and use a low
number of network parameters.
For the generic system additionaly pooling

\subsection{General setup}

The acoustic signal is split into frames. Each frame contains $2048$ samples of the signal, and frames are half-overlapped.
The logarithmic spectral amplitude of each frame is obtained.
Then, the input data is mean-variance normalised, and mean and variance values estimated
on the training data are also applied to the testing data.

The hidden and output nodes use sigmoid and softmax activation functions, respectively.
Stochastic gradient descent is used for network optimisation, and the binary cross-entropy objective function is employed.
The number of epochs is 70 and the minibatch size is set to 10.
The learning rate and momentum parameters are set to 0.01 and 0.9, respectively.
The training data is balanced with regards to classes by randomly selecting samples of non-alarm class.
No unsupervised pre-training of NN is performed. For simplification purposes, the described network configuration is used in all the experiments.

\subsection{Generic input system}

The network structure employed by this system is designed for a \textit{generic} type of alarms,
i.e. is used for all alarm classes. This means that no specific knowledge about
the spectro-temporal characteristics of alarm classes is exploited by this system.

\begin{figure}[h!]
\centerline{\includegraphics[width=0.65\columnwidth]{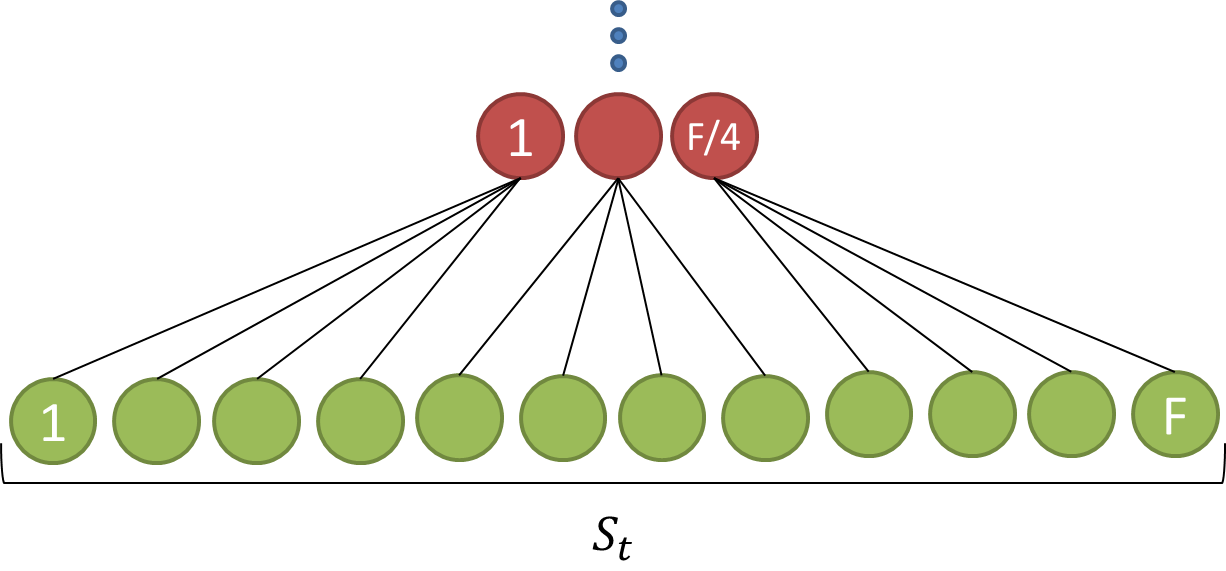}}
\caption{Partially connected hidden layer for weighted averaging in frequency.}
\label{fig:FrequencyPooling}
\end{figure}

In the baseline setup the whole spectral frame is introduced at the input, thus, the input size is 1024.
In our experiments we observed that even using only 8 hidden units the NN is very prone to overfitting.
For that reason, max pooling is used at the input to compress the spectral representation
and reduce the number of parameters to be trained. The max pooling strategy is used with
convolutional neural networks to reduce spectral variance \cite{sainath_improvements_2013}.
It seems to fit well our task as high spectral peaks corresponding to alarms are supposed to be preserved.
In this work, we use either uniform (same structure as in Figure~\ref{fig:FrequencyPooling})
or mel-scale filterbank based (with 60 filters) distribution of pooling filters,
so the input layer size is reduced to either 256 or 60 units, respectively.

Further, we explore the inclusion of partially connected hidden layers, which 
actually are simple unidimensional convolutional layers of one filter with limited weight sharing
and no overlapping.

Figure~\ref{fig:FrequencyPooling} shows a partially connected hidden layer for frequency weighed averaging,
where the weighting filters are uniformly distributed and are non-overlapping.
Here $F$ denotes the number of input spectral bins of frame $S_{t}$.
Since alarms occupy narrow frequency regions, a small (4 units) filter width is used.
Note that no information about the alarm-specific frequencies is provided to the layer.

\begin{figure}[h!]
\centerline{\includegraphics[width=0.65\columnwidth]{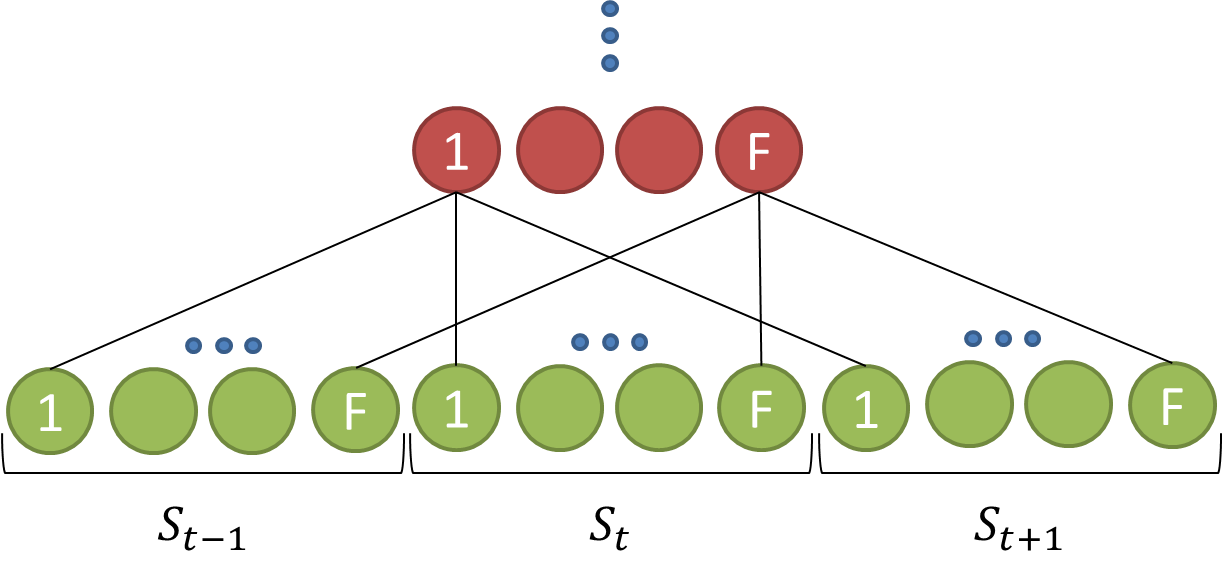}}
\caption{Partially connected hidden layer for weighted averaging in time.}
\label{fig:TemporalPooling}
\end{figure}

Figure~\ref{fig:TemporalPooling} shows a partially connected hidden layer for weighted averaging in time.
A temporal context of several frames is included by this layer, and the smoothed representation of the spectral frame is obtained.

\subsection{Class-specific input system}

In the class-specific system we assume that the particular spectral and temporal properties of alarms are known.
In fact, only the spectral information is used, since the inclusion of the
temporal information (i.e. signal and silence interval duration) would require many more
network weights to be trained (we tried it and the results were clearly worse in our experiments).
Only the logarithmic spectral amplitudes at the alarm-specific frequency bins and a few bins around them are 
employed as input features.
Note that those alarm-specific frequencies include all frequencies from all the versions of the alarm class.
Neither pooling strategies nor partially connected layers for frequency weighted averaging are used in this system.

\subsection{Experimental results}

The development of the two model-based systems was first carried out for the alarm class
that has only one version and the largest number of samples in our database (a8).
In this section we present results for the generic and class-specific input systems for that alarm class
and in terms of Equal Error Rate (EER), which corresponds to both MR and FAR metrics being equal.

Then, the best setups of the generic and class-specific input systems were extended to all the considered alarm classes.
The two systems perform similarly for other classes, with some minor exceptions.
The results over all the alarm classes are presented in Section~\ref{sec:Discussion}
together with the best results obtained by the systems following other approaches.

Table~\ref{tab:NNsetups} summarises the results obtained by the model-based detection systems.
For each experiment, the number of trained NN parameters is provided.
The first row of the table contain the baseline results and the next two rows the results obtained by the generic input system employing
different pooling strategies, i.e. Uniform Max Pooling (UMP) and Mel-Scale Max Pooling (MSMP). 
An additional Fully-Connected (FC) hidden layer of 8 units is used.
It can be seen that none pooling strategy improves the baseline results, although the UMP
provides comparable results with roughly one forth number of parameters.
The results using MSMP are worse (by 6.66\% compared to the baseline results),
most probably due to its stronger information reduction.

\begin{table}[t!]
\small
\caption{Alarm detection performance obtained by the model-based systems for alarm class \textit{a8}}
\label{tab:NNsetups}
\centerline{
\begin{tabular}{Bp{2.3cm}CP}
\hline 
\multirow{3}{*}{Input} & \multicolumn{1}{c}{\multirow{3}{*}{System setup}}
& Evaluation metrics (\%) & \multirow{3}{6em}{\centering Number of parameters} \\
                 & & MR = FAR & \\
\hline \hline
\multirow{5}{1.25cm}{Generic} & FC (baseline) & 23.44 & 8208 \\
\hhline{~---}
& UMP + FC & 23.55 & 2064 \\
& MSMP + FC & 25.00 & 496 \\
\hhline{~---}
& UMP + FW & 18.27 & 384 \\
& \textbf{MSMP + TW} & \textbf{13.87} & 420 \\
\hline
\multirow{2}{1.25cm}{Class- specific} & TW & 13.23 & 202 \\
& \textbf{TW + FC} & \textbf{10.69} & 376 \\
\hline
\end{tabular}}
\end{table}

The third part of Table~\ref{tab:NNsetups} shows the generic input system performance
when partially connected hidden layers are included in the NN structure
for Frequency Weighting (FW) and Time Weighting (TW).
In order to keep the number of trained parameters small, max pooling is applied at the input of the NN.
After trying different combinations, UMP is used before the partially connected hidden layer for weighting in frequency,
while MSMP is used before the partially connected hidden layer for weighting in time,
and in both cases the detection results are improved.
The higher improvement resulting from TW (40.83\%) than from FW (22.06\%) indicates that
the temporal context is more important than the frequencial one. No additional hidden layers are used in these experiments.

Finally, last part of Table~\ref{tab:NNsetups} presents the results for the class-specific input system.
We use $\pm$ 2 bins around the bins corresponding to alarm-specific frequencies,
as it provided better results than using only $\pm$ 1 bin, and
may be more suitable when extending the results to other classes.
Therefore, for the alarm class a8, which has 5 specific frequencies, the input size is equal to 25.
This input is further used with the partially connected hidden layer for weighted averaging in time (TW), which yields 19.18\% relative
improvement in comparison to the network topology with a FC hidden layer.
Notice that the EER is smaller than the best one for the generic input system, thanks to the use of the alarm spectral knowledge.
Moreover, when an extra FC hidden layer of 8 units is added, the detection error drops to 10.69\%.
Note that an extra hidden layer did not improve results for the generic MSMP+TW system setup.

\section{Combined approach}
\label{sec:KnowledgeBased}

This section describes a detection system that employs machine learning techniques and
knowledge about the particular spectral and temporal characteristics of each alarm class,
which are included at different stages.
Each class-specific detector consists of the following blocks:
\begin{enumerate}[itemsep=0pt,partopsep=0pt]
 \item Modelling of the alarm spectral structure, which includes two stages: feature extraction and statistical modelling.
 \item Modelling of the alarm temporal structure, which is based on aggregating the outputs of the statistical models along time.
 \item Post-processing and decision stages. 
\end{enumerate}

\subsection{Modelling of the alarm spectral structure}

The acoustic signal is split into frames of 2048 points, half-overlapped.
The baseline features are 18 frequency-filtered logarithmic filter-bank energies
and their first temporal derivatives (36 features in total), extracted for each frame~\cite{nadeu_ff_2001}.
These are generic features that cover the entire frequency bandwidth.

Since alarms consist only of sinusoidal components, the proposed feature extraction scheme employs detection of sinusoids.
It is performed using an algorithm introduced in~\cite{jancovic_icassp_2011},
which tackles the detection of sinusoidal components as a pattern recognition problem.
Each frame is rectangularly windowed and padded with 2048 zeros to obtain a finer sampled spectrum.
Sinusoid detection is performed independently for each frame.
A given spectral peak $k_p$ is characterised by a feature vector that contains the spectral magnitude shape and phase continuity information around the peak. 
The distribution of that multivariate feature vector is modelled using a 32-component Gaussian mixture.
A model is obtained for spectral peaks corresponding to noise and another one corresponding to sinusoidal signals, at various SNRs.
For a given spectral peak, the log-likelihood is obtained on both the sinusoidal model and on the noise model.
A low log-likelihood value (randomly generated) is assigned to non-peak spectral points.

For each frame, a feature vector is formed by selecting
the log-likelihood values obtained from the sinusoidal detection
in $\pm20$~Hz frequency interval around each alarm-specific frequency.
In each interval, only one sinusoidal and one noise log-likelihoods are chosen,
which correspond to the spectral point with the maximum value of the sinusoidal model likelihood in that interval.
The amplitude structure of the alarms along frequency is incorporated by including in the feature vector the magnitude values of the chosen spectral points.
To disregard the effect of volume, in each frame these magnitudes are normalised by their sum.

The extracted features are further modelled with a Gaussian mixture,
and, for each alarm class, an alarm and a non-alarm models are obtained.
Each model is a single Gaussian probability density function with diagonal covariance matrix
as, in our experiments, this resulted in better detection performance than using more mixture components.

For the best-performing feature extraction setup, we also carried out experiments using pre-trained neural networks \cite{dahl_context-dependent_2012},
where an unsupervised pre-training was performed using a Gaussian-Bernoulli RBM \cite{hinton_practical_2012} and a supervised backpropagation training was performed
after adding a label layer on top of the network, making the model discriminative. A single hidden layer of 32 units was used.
As no performance improvement was obtained, the results of these experiments are not reported.

\subsection{Modelling of the alarm temporal structure}
\label{sec:TemporalModelling}

\begin{figure}[t!]
\centering
\includegraphics[width=95mm, height=55mm]{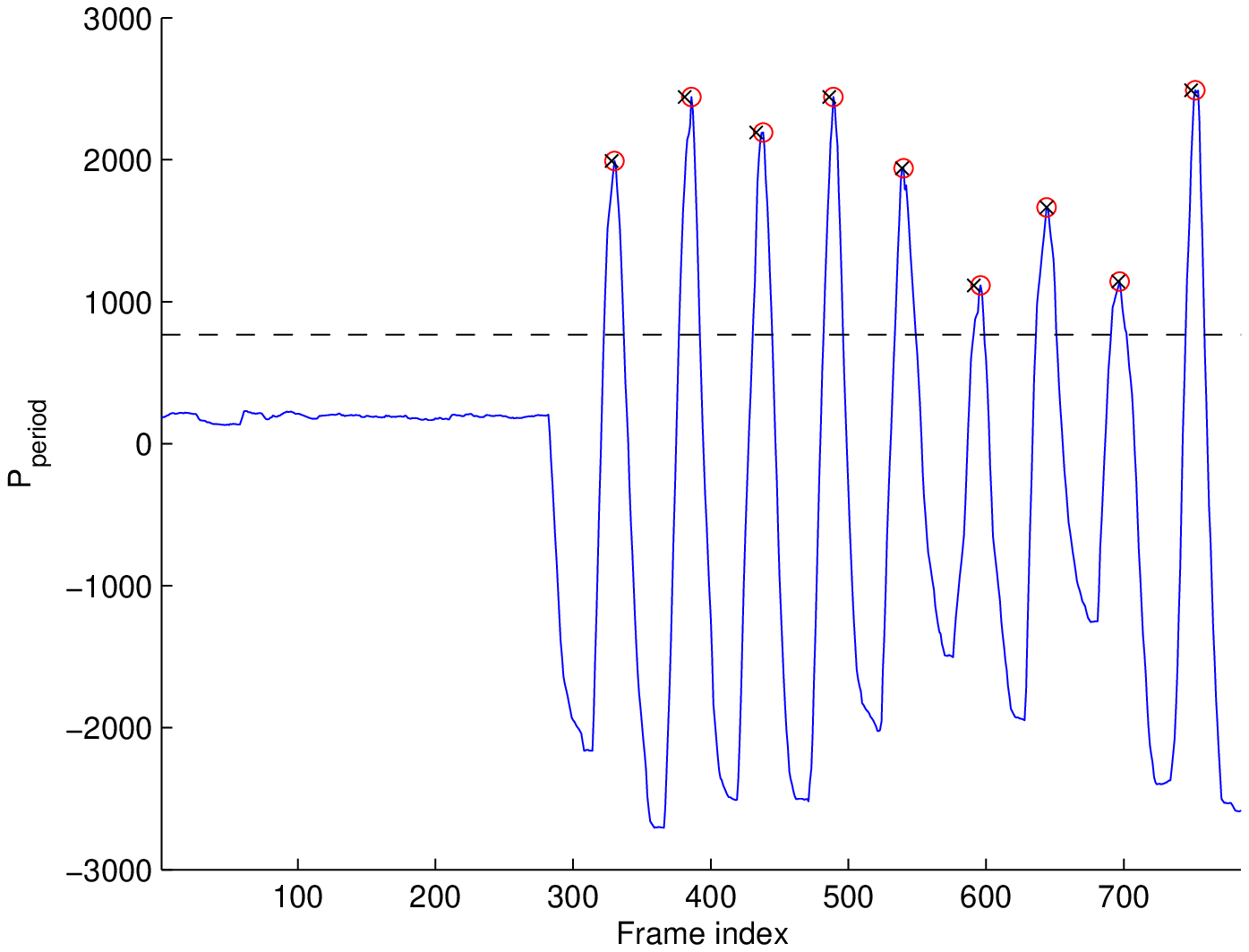}
\caption{The output of the period probability estimation.
Circles correspond to the estimated period timestamps after applying a threshold and crosses are the reference period timestamps.}
\label{fig:P_period}
\end{figure}

The log-posteriors of the alarm and the non-alarm class are calculated for each frame based on the probabilities obtained
from the statistical models - the logarithm is taken after those probabilities are normalised to sum up to one.
The information about the long-term temporal structure of alarms
is incorporated by aggregating these frame-level log-posteriors along
the intervals corresponding to durations of signal and silence segments in every alarm period.
At each frame $t$, the probability of it being the first frame of the alarm period is calculated as
\begin{equation}
  P_{period}(t) = \sum\limits_{i=t}^{t+L_{sig}-1} (P_{A}-P_{NA}) +
  \sum\limits_{i=t+L_{sig}}^{t+L_{sig}+L_{sil}-1} (P_{NA}-P_{A})
\label{Eq:P_period}
\end{equation}
where $P_{A}$ and $P_{NA}$ are log-posteriors of the alarm and non-alarm class, $L_{sig}$ and $L_{sil}$ are, correspondingly, the duration of signal and silence intervals in an alarm period.

An illustration of the output obtained from computing that aggregated probability is given in Figure~\ref{fig:P_period}.
According to the expression defined in (\ref{Eq:P_period}), each peak of the curve corresponds to the first frame of the estimated alarm period.

\subsection{Post-processing and decision}
\label{sec:PostProcessing}

Three alternative post-processing and decision schemes were applied after the modelling blocks.

In the first one, each frame was first classified either as alarm or non-alarm based on the likelihoods obtained from the respective models.
Then, the resulting sequence of labels was smoothed via majority voting.
The length of the smoothing window was set to be the minimum of the signal and silence interval length in an alarm period.

Conversely, in the second one, the decision is taken at the period level.
The period probability $P_{period}(t)$ from (\ref{Eq:P_period}) was subjected to a class-specific thresholding and
the peaks of this probability curve above the threshold were chosen
as the beginning of the detected alarm periods (circles in Figure~\ref{fig:P_period}).
Note that the class-specific threshold was chosen so as to provide the best period-level performance.

Thirdly, a parallel combination of the previous two schemes was applied as follows.
If none of the detected alarm periods obtained from the second scheme coincides with
a sequence of consecutive frames of any length detected from the first scheme as belonging to the alarm class
or is around it with a tolerance of $\pm L_{sig}/2$, the frames of that sequence are assigned to the non-alarm class.

\subsection{Experimental results}
\label{sec:Experiments}

Table~\ref{tab:PostProcessing} shows the results for the detection system based on the combined approach.
The first part of the table presents the results obtained when
no post-processing is incorporated (therefore, the reported MR and FAR metric scores correspond to the EER value),
and either the baseline generic features or the features based on sinusoid detection are used.
It can be seen that the features based on sinusoid detection, which exploit the knowledge of
alarm properties, can significantly outperform (62.12\%) the baseline generic features.
The former features are used in posterior experiments.

The second part of the table shows the experimental results
when temporal modelling and smoothing are incorporated, as described in Section~\ref{sec:PostProcessing}.
Note that all these results correspond to the same EER value, i.e. 13.37\%.
It can be seen that none post-processing scheme improves MR metric scores compared to not performing any post-processing at all,
but all schemes improve the FAR metric scores to a large extent (up to 87.8\% relative improvement in the best case).
Moreover, all the post-processing schemes are able to improve the PB-ERR scores.

\begin{table}[t!]
\small
\caption{Alarm detection performance obtained by the combined system}
\label{tab:PostProcessing}
\vspace{2mm}
\centerline{
\begin{tabular}{p{4.35cm}ccc}
\hline 
\multicolumn{1}{c}{\multirow{2}{*}{Post-processing}} & \multicolumn{3}{c}{Evaluation metrics (\%)} \\ 
                 & MR  & FAR  & PB-ERR \\
\hline \hline
None (baseline)        & 35.30 & 35.30 & 93.45 \\
None (sinusoid detection) & \textbf{13.37} & 13.37 & 68.97 \\
\hline
Smoothing (S)        & 13.70 & 9.61 & 53.62 \\
Temporal modelling (TM)    & 33.56 & 2.36 & 36.27 \\
Combination (S \& TM) & 32.49 & \textbf{1.57} & \textbf{33.10} \\  
\hline 
\end{tabular}}
\end{table}

In general, smoothing provides better results at the frame level, while temporal modelling performs
better at the period level. It can be seen that smoothing slightly increases MR, but is able to significantly improve results in terms of FAR
(which corresponds to -2.64\% and 28.12\% relative improvement).
Temporal modelling, on the other hand, reduces even more strongly the FAR error (by 82.35\%, relatively) and is not performing
well in terms of MR metric, but gives better period-level score.
Although there is a big difference between frame-level metrics in percentage,
in terms of absolute number of frame errors the deterioration of MR results is smaller than the improvement of FAR results.
The best PB-ERR metric score corresponds to the combination of both smoothing and temporal modelling (\textit{S} \& \textit{TM}),
which is 52\% relatively better than the baseline not using any post-processing.
Moreover, it should be noted that the combination of both schemes outperforms the temporal modelling not only in terms of PB-ERR,
but also at the frame level.

\section{Discussion}
\label{sec:Discussion}

The non-model-based detection system that uses matched filter and morphological operators was presented first.
The system requires that at least a sample of each target alarm is available for each of its alternative versions.
That sample is used as a reference signal at the matched filter stage, so it should be clean enough.
In order to enhance it, the system includes a prior filtering step, which
suppresses the content at all frequencies not relevant to the alarm class.
As usual for non-model-based systems, the performance depends greatly on the proper choice of the decision threshold.
Note that the threshold estimation is based on the non-alarm training data.

Further, a model-based approach employing neural networks was explored.
Following this approach, two detection systems were developed:
1) a generic input system where no specific knowledge about the alarm properties is used to construct the network;
2) a class-specific input system where the information
about the particular spectral properties of alarms is incorporated at the input.
With the experiments it has been shown that the class-specific input system clearly outperforms the generic input one.
Nevertheless, while the class-specific system has to be adapted to each alarm class,
the generic input system has the advantage of using the same network topology for all the classes,
so it can be extended to new alarm classes in an easier way.
Due to the limited amount of data available, the number of network parameters to train must be constrained.
For that reason, two types of partially connected layers that weight the input data in either time or
frequency have been considered. It has been shown that they provide better detection results
than the fully connected layers, besides reducing the complexity of network training.
Moreover, for both generic and class-specific input systems,
the layer exploiting temporal context improved the results to a larger extent than the one which does it in frequency.
Additionally, it should be noted that, according to the experimental results,
there is not a clear rule relating the number of network parameters and the detection performance.

The combined system strongly relies on the feature extraction process, and
is based on exploiting the knowledge about the particular spectro-temporal properties of alarms.
First, the spectral information about alarms is captured at the feature level.
The best-performing features are based on the output of a sinusoidal detector
complemented by the amplitude structure information at the alarm-specific frequency regions.
Second, after using the statistical models of those features, the temporal information is included at the post-processing step.
In particular, the period probability estimate is obtained at each frame by aggregating the log-posterior probabilities
from statistical models along the signal and silence intervals in the alarm period.
It has been shown that the detection system benefits largely from the introduction of
both spectral and temporal information, and both were important to improve the detection performance.
Note that the inclusion of the temporal modelling provides more than 33\% improvement in absolute terms at the alarm period level,
suggesting again that the temporal information is important for acoustic alarm detection.

Table~\ref{tab:SystemsResults} summarises the results obtained by the proposed detection systems over all alarm classes.
For the model-based Generic Input (GI) and the Class-Specific Input (CSI) systems
the results that correspond to their best performance versions
are provided, i.e. \textit{MSMP + TW} and \textit{TW + FC},
respectively (see Table~\ref{tab:NNsetups}).
The EER obtained before the post-processing stage is reported and
either smoothing~(S) or temporal modelling (TM) post-processing is applied in the systems that classify at the frame level.
Note that, for comparison purposes, smoothing is also applied to the two model-based systems,
even the generic input one, despite that for setting the smoothing window length,
the information about the durations of signal and silence intervals in the alarm period are used.

\begin{table}[!t]
\small
\caption{Alarm detection performance obtained by the developed systems for all alarms}
\label{tab:SystemsResults}
\centerline{
\begin{tabular}{p{1.3cm}p{2.9cm}cccc}
\hline 
\multirow{2}{1.3cm}{\centering Detection level} &\multicolumn{1}{c}{\multirow{2}{*}{System}} & \multicolumn{1}{c}{\multirow{2}{*}{EER}} &
\multicolumn{3}{c}{Evaluation metrics (\%)} \\ 
                 & & & MR  & FAR & PB-ERR \\
\hline \hline
\multirow{3}{*}{F (frame)} & GI model + S & 17.76 & 33.52 & 7.94 & 54.80 \\
& CSI model + S & \textbf{11.13} & 17.72 & 5.22 & 53.75 \\
& Combined + S & 13.37 & 13.70 & 9.61 & 53.62 \\
\hline
\multirow{2}{*}{P (period)} & Combined + TM & 13.37 & 33.56 & 2.36 & 36.27 \\
& Non-model & -- & 42.38 & 0.53 & 34.24 \\
\hline
F \& P & Combined + S\&TM & 13.37 & 32.49 & 1.57 & \textbf{33.10} \\
\hline
\end{tabular}}
\end{table}

As can be expected, the systems performing detection at the frame level provide better overall performance in terms of
the frame-level metrics, and, similarly, the systems operating at the period level provide
better results in terms of the period-level metric. Also, it can be seen that the inclusion of knowledge
about the alarm properties is beneficial for the detection performance. In fact, going from top to bottom in
Table~\ref{tab:SystemsResults}, the systems use more and more knowledge and the results are better and better.
In the middle, the combined approach provides a kind of balance between the frame-level and the period-level performances.
And the smallest error percentage at the meaningful alarm period level is achieved by
the machine learning system that uses the parallel combination of smoothing and temporal modelling,
and thus combines the detection at the frame and at the alarm period levels.
This suggests that other combinations between systems operating at different levels may bring further reduction in error rates.
Notice that the detection errors obtained by the proposed systems are rather high; this can be attributed
to two facts: the real-world NICU environment is very noisy and the available dataset is small.

To get a further insight into the performance of the systems,
a more detailed analysis is carried out in terms of errors by alarm class (Table~\ref{tab:PBResults}, where the 
absolute number of false alarm (FA) and miss (M) errors is reported along with the corresponding PB-ERR score)
and the confusion between the classes (Tables~\ref{tab:ConfusionMatrix_SP} and~\ref{tab:ConfusionMatrix_KTM}).
We report the results at the period level due to its importance for the medical application,
and for both period-oriented systems: the non-model based system
and the combined system that includes smoothing and temporal modelling.

\begin{table*}[!t]
\small
\caption{Class-based scores obtained by the systems at the period level}
\label{tab:PBResults}
\centerline{
\begin{tabular}{p{2.8cm}Cccccccc}
\hline 
\multicolumn{1}{c}{\multirow{2}{*}{System}} & \multirow{2}{6em}{\centering Evaluation metric}
  & \multicolumn{7}{c}{Alarm class} \\
 & & a1 & a3 & a6 & a7 & a8 & a10 & a16 \\
\hline \hline
\multirow{3}{*}{Non-model} & FA & \cellcolor{grey}{\textbf{99}} & \cellcolor{lightergrey}{18} & \cellcolor{lightgrey}{48} 
& \cellcolor{lightgrey}{23} & \cellcolor{lightergrey}{37} & \cellcolor{lightergrey}{13} & \cellcolor{grey}{\textbf{76}} \\
& M & \cellcolor{darkergrey}{\textbf{164}} & \cellcolor{lightergrey}{19} & \cellcolor{lightergrey}{28} & \cellcolor{lightgrey}{30}
& \cellcolor{darkergrey}{\textbf{296}} & \cellcolor{lightgrey}{22} & \cellcolor{grey}{\textbf{48}} \\
\hhline{~--------}
& PB-ERR (\%) & 63.99 & \textbf{14.29} & \textbf{17.84} & 23.98 & 53.11 & 24.82 & \textbf{41.61} \\
\hline
\multirow{3}{*}{Combined + S\&TM} & FA & \cellcolor{lightgrey}{70} & \cellcolor{lightergrey}{19} 
& \cellcolor{darkergrey}{\textbf{223}} & \cellcolor{lightgrey}{33} & \cellcolor{lightgrey}{102} & \cellcolor{lightergrey}{8} & \cellcolor{darkergrey}{\textbf{369}} \\
& M & \cellcolor{grey}{\textbf{84}} & \cellcolor{grey}{\textbf{46}} & \cellcolor{lightergrey}{16} 
& \cellcolor{lightergrey}{15} & \cellcolor{grey}{\textbf{171}} & \cellcolor{lightergrey}{5} & \cellcolor{lightgrey}{46} \\
\hhline{~--------}
& PB-ERR (\%) & \textbf{33.33} & 27.90 & 38.99 & \textbf{19.51} & \textbf{33.50} & \textbf{8.50} & 69.98 \\
\hline
\end{tabular}}
\end{table*}

For the confusion analysis, all the testing samples of each class are used as input of 
each binary detector, and the \textit{i}\textsuperscript{th} column
in the tables shows the percentages of samples of each actual class that get a positive detection by the
\textit{i}\textsuperscript{th} detector.
The  evaluation parameter $T_{tol}$ is set
to $5\%$ of alarm period duration to reduce possible overlaps between tolerance intervals of different alarms.
The rows are normalised by the total number of samples of the corresponding alarm class.
The diagonal elements in italic correspond to correct period detections of the alarm class, and
the elements in bold denote the major sources of confusion between the alarm classes ($>$5\%).
Note that, due to the confusion calculation routine, the matrix is not symmetric
and the rows don't sum up to 100\% as some alarms may not be detected by any class-specific detector,
while others may be detected by more than one. Also, the temporal overlaps between the alarm classes were taken into account.

For explaining the results, we consider the following properties of the alarm classes that may influence the detection performance:
1) time domain structure (i.e. duration of the signal and silence intervals, time concatenation of different frequencies); 
2) frequency domain structure (i.e. fundamentals and harmonics, frequency sharing between classes); 
and 3) presence of different versions of the same alarm (i.e. number, spectro-temporal properties).

It can be seen from Table~\ref{tab:PBResults} that for both systems there is a strong dispersion along class-specific detectors in terms of error rates.
The best results are obtained for the class a3 by the non-model-based system and for the class a10 by the combined system.
This is in line with the fact that for these systems the uniqueness and richness of, correspondingly, time structure and frequency content are important.
In this case, unlike other alarms, a3 has two consecutive tones in the signal interval, and 
a10 has the most distinctive set of frequency components among all the considered alarms.

The combined system provides better results for a1, a7 and a10, i.e. for all alarm classes that have several versions except a3, 
which, as was mentioned above, has a richer time structure that is better exploited by the non-model-based system.
In the combined system, the frequency order is discarded and the features correspond to all the alarm-specific frequencies, including all the versions,
resulting in a larger feature vector which is beneficial for the detection performance.

Overall, in terms of absolute values, the non-model-based system produces more miss errors while the combined system produces more false alarm errors,
which mainly reflects how each system deals with similarities between the alarm classes.

Poor detection results are obtained for the classes a1, a8 and a16. As was mentioned before, a16 shares spectro-temporal properties with a1 and a8,
but, unlike these two classes, it has only one frequency component, which coincides with the fundamental frequencies of both a1 and a8.
Also, the frequency content of a8 is very similar to that of both versions of a1.

The similarities between the classes a1, a8 and a16 are also reflected in high confusion values 
for both the non-model-based system (Table~\ref{tab:ConfusionMatrix_SP}) and the combined system (Table~\ref{tab:ConfusionMatrix_KTM}).
Apart from that, there is confusion between the classes a6 and a16 in the combined system,
which may be due to the fact that both alarms have only one specific frequency and are similar in the time domain.
The high confusion between a3 and a16 for the same system might be explained by the similar frequency content of a16 and one of the versions of a3.

\begin{table}[!t]
\caption{Information about confusion at the period level for the non-model based system (in \%)}
\label{tab:ConfusionMatrix_SP}
\small
\centerline{
\begin{tabular}{cc | c ccc ccc}
\hline 
\multicolumn{9}{c}{Class-specific detector \smallskip} \\
\multirow{9}{*}{\rotatebox[origin=c]{90}{Actual class}} & & a1 & a3 & a6 & a7 & a8 & a10 & a16 \\
 \hline
& a1 & \textit{27.7} & 0 & \cellcolor{lightergrey}{0.4} & \cellcolor{lightergrey}{0.4} & \cellcolor{lightgrey}{2.1} & 0 & 0 \\
& a3 & \cellcolor{lightgrey}{1.5} & \textit{73.1} & \cellcolor{lightergrey}{0.8}
  & \cellcolor{lightgrey}{1.5} & \cellcolor{grey}{5.4} & \cellcolor{lightgrey}{1.5} & 0 \\
& a6 & \cellcolor{lightgrey}{2.0} & \cellcolor{lightergrey}{0.5} & \textit{8.9} & 0 & \cellcolor{lightergrey}{0.5}
  & \cellcolor{lightergrey}{0.5} & \cellcolor{lightergrey}{0.5} \\
& a7 & \cellcolor{lightergrey}{0.9} & \cellcolor{lightergrey}{0.9} & \cellcolor{lightergrey}{0.9} & \textit{21.9} & \cellcolor{lightgrey}{2.6} & 0 & 0 \\
& a8 & \cellcolor{lightgrey}{2.9} & 0 & 0 & 0 & \textit{32.3} & 0 & 0 \\
& a10 & \cellcolor{lightgrey}{1.3} & 0 & \cellcolor{lightgrey}{4.0} & 0 & 0 & \textit{41.3} & 0 \\
& a16 & \cellcolor{darkergrey}{\textbf{11.1}} & \cellcolor{lightergrey}{0.7} & 0 & \cellcolor{lightgrey}{1.5}
  & \cellcolor{darkergrey}{\textbf{28.9}} & 0 & \textit{60.7} \\
\hline
\end{tabular}}
\end{table}

\begin{table}[!t]
\caption{Information about confusion at the period level for the combined system (in \%)}
\label{tab:ConfusionMatrix_KTM}
\small
\centerline{
\begin{tabular}{cc | c ccc ccc}
\hline 
\multicolumn{9}{c}{Class-specific detector \smallskip} \\
\multirow{9}{*}{\rotatebox[origin=c]{90}{Actual class}} & & a1 & a3 & a6 & a7 & a8 & a10 & a16 \\
 \hline
& a1 & \textit{54.2} & \cellcolor{lightgrey}{2.9} & \cellcolor{lightgrey}{2.5} & 0 & \cellcolor{darkergrey}{\textbf{20.6}}
  & \cellcolor{lightergrey}{0.4} & \cellcolor{lightergrey}{0.8} \\
& a3 & \cellcolor{lightgrey}{1.5} & \textit{52.3} & \cellcolor{lightgrey}{4.6} & \cellcolor{lightergrey}{0.8}
& \cellcolor{lightgrey}{2.3} & 0 & \cellcolor{lightergrey}{0.8} \\
& a6 & \cellcolor{grey}{6.9} & 0 & \textit{11.3} & 0 & \cellcolor{lightergrey}{0.5} & 0 & 0 \\
& a7 & \cellcolor{lightgrey}{1.8} & 0 & \cellcolor{lightgrey}{1.8} & \textit{48.3} & 0 & 0 & 0 \\
& a8 & \cellcolor{grey}{\textbf{10.0}} & \cellcolor{lightergrey}{0.9} & 0 & 0 & \textit{57.1}
& 0 & \cellcolor{lightergrey}{0.9} \\
& a10 & \cellcolor{lightgrey}{1.3} & 0 & \cellcolor{lightgrey}{1.3} & 0 & 0 & \textit{72.0} & 0 \\
& a16 & \cellcolor{darkergrey}{\textbf{48.9}} & \cellcolor{darkergrey}{\textbf{19.3}} & \cellcolor{darkergrey}{\textbf{11.1}}
  & 0 & \cellcolor{darkergrey}{\textbf{132.6}} & \cellcolor{lightergrey}{0.7} & \textit{57.8} \\
\hline
\end{tabular}}
\end{table}

\section{Conclusions}

In this paper, our work on the challenging problem of automatic acoustic alarm detection in a
NICU environment was reported. Several detection systems were proposed for the detection of particular
types of alarms, which are based on the approaches that deal with the problem from different
perspectives. The detection errors obtained by the proposed systems are rather high, presumably due
to both the fact that the real-world NICU environment is noisy and to the scarcity of available annotated data.

In terms of the medical application, the best results were obtained by the system that combines: 
1) non-model-based and model-based approaches, and 
2) frame-level and period-level detection.
As can be expected, all the detection systems got the worst metric scores for the alarm classes that share similar spectro-temporal properties.
In future work, the apparent complementarity among the various detection systems can be exploited
by using the detector from one approach or the other depending on the type of alarm
or combining the decisions from the systems in posterior stage.

\begin{backmatter}

\section*{Competing interests}
  The authors declare that they have no competing interests.

\section*{Author's contributions}
This paper presents results from the doctoral research of GR under the supervision of CN. 
The non-model-based and model-based systems were developed in the scope of the final 
degree project and master thesis of SGQ and APL, respectively, under the co-supervision of 
GR and CN. All authors read and approved the final manuscript.

\section*{Acknowledgements}
  This work has been supported in part by the Catalan Government (grant FI-DGR, 2012-2015), the Spanish government
(contracts TEC2010-21040-C02-01, TEC2012-38939-C03-02 and TEC2015-69266-P)
and the European Regional Development Fund (ERDF/FEDER, 2015).
The authors are grateful to Ana Riverola de Veciana and Blanca Mu\~{n}oz Mahamud for
their work on the database collection and on the medical aspects of this study,
and to Peter Jan\u{c}ovi\u{c} and to M\"{u}nevver K\"{o}k\"{u}er
mainly for their contributions to the feature extraction of the combined system, and to
Vanessa Sancho Torrents and to Francisco Alarc\'{o}n Sanz for their work on the database annotation.


\bibliographystyle{bmc-mathphys} 
\bibliography{Paper_SC_v.2}      

\end{backmatter}
\end{document}